\documentclass[aps,prd,preprint,groupedaddress]{revtex4-1}
\usepackage[utf8]{inputenc}
\usepackage[english]{babel}
\usepackage[letterpaper,margin=1in]{geometry}

\usepackage[dvipsnames]{xcolor}
\usepackage{amsmath,amssymb,amsfonts,enumerate,float,graphicx,mathrsfs,mathtools,fancyvrb,slashed,pdfpages,empheq,tabularx,caption,multirow,subcaption}

\usepackage[inline,shortlabels]{enumitem}

\allowdisplaybreaks

\usepackage{colortbl}
\usepackage{tikz,pgfplots}

\usetikzlibrary{calc,patterns,decorations.pathmorphing,decorations.markings,arrows}

\graphicspath{{./figures/}}

\usepackage[most]{tcolorbox}

\tcbset{
	frame code={}
	center title,
	left=0pt,
	right=0pt,
	top=0pt,
	bottom=0pt,
	colback=yellow!25,
	colframe=white,
	width=\dimexpr\textwidth\relax,
	enlarge left by=0mm,
	boxsep=5pt,
	arc=0pt,outer arc=0pt,
}

\newcommand\myshade{85}
\usepackage{hyperref}
\hypersetup{
	linkcolor  = violet!\myshade!black,
	citecolor  = YellowOrange!\myshade!black,
	urlcolor   = Aquamarine!\myshade!black,
	colorlinks = true,
}

\newcommand{\pp}[1]{\left ( #1 \right )}
\newcommand{\bb}[1]{\left [ #1 \right ]}
\newcommand{\cc}[1]{\left \{ #1 \right \}}

\newcommand{\del}{\partial}

\makeatletter
\newcommand\incircbin{\mathpalette\@incircbin}
\newcommand\@incircbin[2]{\mathbin{\ooalign{\hidewidth$#1#2$\hidewidth\crcr$#1\bigcirc$}}}

\makeatother

\DeclareMathOperator{\tr}{\mathrm{tr}}

\newcommand{\prg}{\\ $  $ \\  }

\newcommand{\nn}{\nonumber\\ &}

\newcommand{\doublevec}[1]{\overset{\text{\scriptsize$\leftrightarrow$}}{#1}}

\def\ba{\begin{eqnarray}}
\def\ea{\end{eqnarray}}
\def\bt{\begin{eqnarray*}}
	\def\et{\end{eqnarray*}}

\makeatletter
\newcommand{\pushright}[1]{\ifmeasuring@#1\else\omit\hfill$\displaystyle#1$\fi\ignorespaces}
\newcommand{\pushleft}[1]{\ifmeasuring@#1\else\omit$\displaystyle#1$\hfill\fi\ignorespaces}
\makeatother

\makeatletter
\AtBeginDocument{\let\LS@rot\@undefined}
\makeatother

\begin{document}
\title{Gravitational form-factors of the $ \pi $ and $ K $ mesons in QCD sum rules}

\author{T. M. Aliev}
\email[]{taliev@metu.edu.tr}
\affiliation{Physics Department, Middle East Technical University, Ankara 06800, Turkey}

\author{K. \c Sim\c sek}
\email[]{ksimsek@u.northwestern.edu}
\affiliation{Department of Physics \& Astronomy, Northwestern University, Evanston, IL 60208, USA}

\date{September 10, 2020}

\begin{abstract}
	Taking into account only the quark part of the energy-momentum tensor, the gravitational form-factors of the $ \pi $ and $ K $ mesons are calculated within the light-cone sum rules method using the distribution amplitudes of these pseudoscalar mesons. The $ Q^2 $-dependence of the relevant form-factors are obtained. Moreover, the mean square mass radii of $ \pi $ and $ K $ mesons are calculated and we find that $ \sqrt{\langle r _{\pi}^2 \rangle} = 0.31 {\rm\ fm} $ and $ \sqrt{\langle r _{K}^2 \rangle} = 0.24 {\rm\ fm} $, respectively. We compare our results for gravitational form-factors with the existing literature results. We obtain that our results on $ \sqrt{\langle r _\pi^2 \rangle} $ are in good agreement with the predictions of NJL model, AdS/QCD model, and other models as well as with the existing experimental data.
\end{abstract}
\maketitle 
\section{Introduction}\label{sec:1}
The energy-momentum tensor (EMT) of hadrons became a more popular object for a deeper understanding of the structure of the hadrons \cite{ref:1,ref:2}. The gravitational form-factors (GFFs) are defined through the matrix element of the EMT between hadronic states \cite{ref:3,ref:4}. The GFFs have been defined for spin-0 and spin-1 mesons and spin-1/2 baryons in many works (see for example  \cite{ref:5,ref:6,ref:7,ref:8,ref:9,ref:10,ref:15,ref:16}.) The GFFs of spin-1 particles have been comprehensively studied in a lot of works \cite{ref:6,ref:7,ref:8,ref:9,ref:10,ref:10.1,ref:11}. One should remark that the study of the GFFs for hadrons began with \cite{ref:14}.
\prg
The GFFs of the $ \pi $ meson were studied in the chiral limit of a hard-wall AdS/QCD \cite{ref:11}, in the NJL model \cite{ref:10}, by using the generalized distribution amplitudes \cite{ref:a2}, and in the lattice QCD \cite{ref:12}. The GFFs of the $ K $ meson were studied in the light-cone quark model \cite{ref:13}. 
\prg 
In this work, we investigate the GFFs of the $ \pi $ and $ K $ mesons within the method of light-cone sum rules (LCSR). In Section \ref{sec:2}, we derive the sum rules for the aforementioned GFFs. In Section \ref{sec:3}, we perform the numerical analysis for the sum rules obtained in the previous section. In Section \ref{sec:4}, we conclude. 
\section{LCSR for the gravitational form-factors of the light pseudoscalar mesons}\label{sec:2}
We start the construction of the LCSR for the calculation of the light pseudoscalar mesons by introducing the following correlation function:
\begin{align}
	\Pi _{\mu\nu\lambda} &= i \int d^4x \ e^{iq\cdot x} \langle \mathcal P (p) \vert T \{T _{\mu\nu}^q (x) j _\lambda (0) \} \vert 0 \rangle 
\end{align}
where $ \mathcal P $ is a pseudoscalar meson of momentum $ p $, $ j _\lambda (0) = \bar u \gamma _\lambda \gamma _5 d $ is its interpolating current, and $ T _{\mu\nu}^q (x) $ is the EMT including only the contribution of the quark fields, given by
\begin{align}
	T _{\mu\nu}^q &= \frac i4 \bb{
		\bar \psi \pp{\doublevec{\mathcal D} _\mu \gamma _\nu + \doublevec{\mathcal D} _\nu \gamma _\mu} \psi -
		g _{\mu\nu} \frac i2 (\doublevec{\mathcal D} - m _q) \psi
	}
\end{align}
where we have defined $ \doublevec{\mathcal D} _\mu := \doublevec{\del} _\mu \pm i g A _\mu^a \frac{\lambda ^a}{2} $. The term in the EMT proportional to the metric can be rewritten as $ g _{\mu\nu} (\doublevec{\slashed{\mathcal D}} - m _q) \psi \approx g _{\mu\nu} (1+\gamma _m) m _q \bar q q $ where $ \gamma _m $ denotes the anomalous dimension of the mass operator. We are considering the chiral limit in this work, namely $ m _q \to 0 $, and thus the second term in the EMT can be safely ignored. 
\prg 
Let us start our analysis by calculating the correlation function from the phenomenological side. We obtain the representation of the correlation function in the hadronic side by inserting a complete set of mesons carrying the same quantum numbers as the light pseudoscalar meson $ \mathcal P $ and isolating the contribution of the ground state as follows:
\begin{align}
	\Pi _{\mu\nu\lambda} &= \frac{\langle \mathcal P (p) \vert T _{\mu\nu}^q \vert \mathcal P (p') \rangle \langle \mathcal P (p') \vert j _\lambda \vert 0 \rangle}{{p'}^2 - m _{\mathcal P}^2} + \cdots \label{3}
\end{align}
where $ \cdots $ indicates the contributions from higher states and the continuum. The matrix elements in Eq. \eqref{3} are given as
\begin{align}
	\langle \mathcal P (p') \vert j _\lambda \vert 0 \rangle = f _{\mathcal P}  p' _\lambda \label{4}
\end{align}
where $ f _{\mathcal P} $ is the decay constant of the pseudoscalar meson $ \mathcal P $ and $ p' _\lambda $ is its 4-momentum. The most general form of the matrix element between spinless particles can be written in the following way:
\begin{align}
	\langle \mathcal P (p) \vert T _{\mu\nu}^a \vert \mathcal P (p') \rangle &= 2P _\mu P _\nu F _{0}^a (q^2) +\frac 12 (q _\mu q _\nu - q^2 g _{\mu\nu}) F _{1}^a(q^2) + 2 m _{\mathcal P} ^2 F _{2}^a(q^2) g _{\mu\nu} \label{5}
\end{align}
where $ P = \frac 12 (p+p') $ and $ q=p'-p $. The superscript $ a $ in Eq. \eqref{5} means quarks or gluons. In the present work, we restrict ourselves by considering only the quark part of the EMT. The first two form-factors are individually EMT-conserving and the last one is non-conserving. The conservation of the EMT leads to the following constraints:
\begin{align}
	\sum _{a=q,g} F _{2}^a(q^2) &= 0\\
	\sum _{a=q,g} F _{0}^a(q^2) &= 1 \label{7}
\end{align}
Another constraint follows from the low-energy pion theorem \cite{ref:a4,ref:a5,ref:a6}:
\begin{align}
	\lim _{m _\pi \to 0} \sum _{a=q,g} F _{1}^a(q^2) =-1 \label{8}
\end{align}  
Using Eqs. \eqref{4} and \eqref{5}, from Eq. \eqref{3} for the phenomenological part of the correlation function, we get
\begin{align}
	\Pi _{\mu\nu\lambda} &= \frac{f _{\mathcal P}}{p'^2-m _{\mathcal P}^2} \Big[
		2 F _0^a P _\mu P _\nu P _\lambda 
		+ \frac 12 F _1^a q _\mu q _\nu P_\lambda 
		+ \mbox{other structures}
	\Big] 
\end{align}
Next, we move on to the computation of the correlation function from the theoretical side. Using the explicit forms of the interpolating currents $ j _\lambda $ and $ T _{\mu\nu} ^q $ and making use of the Wick theorem, we obtain
\begin{align}
	\Pi _{\mu\nu\lambda} &= - \frac 14 \int d^4x \ e^{iq\cdot x} \langle \mathcal P (p') \vert 
		\bar u (x) \gamma _\mu \doublevec{\mathcal D} _\nu S (x) \gamma _\lambda \gamma _5 d (0) 
		+ \bar u (0) \gamma _\lambda \gamma _5 \doublevec{\mathcal D} _\mu S (x) \gamma _\nu d (x) 
		\nn + \bar u (x) \gamma _\nu \doublevec{\mathcal D} _\mu S (x) \gamma _\lambda \gamma _5 d (0) 
		+ \bar u (0) \gamma _\lambda \gamma _5 \doublevec{\mathcal D} _\nu S(x) \gamma _\mu d(x) \vert 0 \rangle   
\end{align}
Here, one can see that the light quark propagator in the presence of an external background field is needed, which is given by \cite{ref:17}
\begin{align}
	S (x) &= \frac{i\slashed x}{2\pi^2 x^4} - \frac{ig _s}{16\pi^2 x^2} \int _0^1 du\ [\bar u \sigma _{\alpha\beta} \slashed x + u \slashed x \sigma _{\alpha\beta} ] G ^{\alpha\beta} - \frac{ie _q}{16\pi^2 x^2} \int _0^1 du\ [\bar u \sigma _{\alpha\beta} \slashed x + u \slashed x \sigma _{\alpha\beta}] F ^{\alpha\beta} 
\end{align}
where $ G ^{\alpha\beta} $ and $ F^{\alpha\beta} $ are the gluon and photon field strength tensors, respectively. Performing the calculations involved for the correlation function from the QCD side, we obtain
\begin{align}
	\Pi _{\mu\nu\lambda} &= - \frac 14 \Big\{ 
		\frac{i}{2\pi^2} \sum _i \int du \ \Big[ 
			\frac 14 \langle \mathcal P \vert \bar u \Gamma _i d \vert 0 \rangle \tr \cc{\Gamma _i \gamma _\nu \pp{\frac{\gamma _\mu}{x^4} - \frac{4x _\mu \slashed x}{x^6}} \gamma _\lambda \gamma _5}
			\nn + \frac 14 \langle \mathcal P \vert \bar u \Gamma _i d \vert 0 \rangle \tr \cc{\Gamma _i \gamma _\lambda \gamma _5 \pp{\frac{\gamma _\mu}{x^4} - \frac{4x _\mu \slashed x}{x^6}} \gamma _\nu} 
			+ (\mu \leftrightarrow \nu) 
		\Big]
		\nn - \frac{ig _s}{16\pi^2} \sum _i \int du \ \Big[
			\frac 14 \langle \mathcal P \vert \bar u \Gamma _i G ^{\alpha\beta} d \vert 0 \rangle \nn\times \tr \cc{\Gamma _i \gamma _\lambda \gamma _5 \bb{\bar u \pp{\frac{\gamma _\mu}{x^2} - \frac{2x_\mu \slashed x}{x^4}} \sigma _{\alpha\beta} + u \sigma _{\alpha\beta} \pp{\frac{x _\mu}{x^2} - \frac{2x _\mu \slashed x}{x^4}} } \gamma _\nu }
			\nn + \frac 14 \langle \mathcal P \vert \bar u \Gamma _i G ^{\alpha\beta} d \vert 0 \rangle \tr \cc{\Gamma _i \gamma _\nu \bb{\bar u \pp{\frac{x _\mu}{x^2} - \frac{2x _\mu \slashed x}{x^4}} \sigma _{\alpha\beta} + u \sigma _{\alpha\beta} \pp{\frac{\gamma _\mu}{x^2} - \frac{2x _\mu \slashed x}{x^4}} } \gamma _\lambda \gamma _5} 
			\nn + (\mu \leftrightarrow \nu ) 
		\Big]
	\Big\} \label{9}
\end{align}
In Eq. \eqref{9}, the $ \{ \Gamma _i \} $ is the complete set of Dirac matrices,
\begin{align}
	\Gamma _1 = 1,\ \Gamma _2 = \gamma _5, \ \Gamma _3 = \gamma _\mu,\ \Gamma _4 = i \gamma _\mu \gamma _5,\ \Gamma _5 = \frac{1}{\sqrt 2} \sigma _{\mu\nu}
\end{align}
The matrix elements $ \langle \mathcal P \vert \bar u \Gamma _i d \vert 0 \rangle $ and $ \langle \mathcal P \vert \bar u \Gamma _i G _{\alpha\beta} d \vert 0 \rangle $ are given in terms of the pseudoscalar meson distribution amplitudes (DAs) of different twists. In the LCSR, these DAs are the primary non-perturbative parameters \cite{ref:18,ref:19}:
\begin{align}
	\langle \mathcal P(p) \vert \bar u (x) \gamma _5 d (0) \vert 0 \rangle &= - i \mu _{\mathcal P} \int _0^1 du \ e^{i\bar u p \cdot x}\phi _P (u) \label{11}\\
	\langle \mathcal P (p) \vert \bar u (x) \gamma _\varphi \gamma _5 d (0) \vert 0 \rangle &= -i f _{\mathcal P} p _\varphi \int _0^1 du\ e^{i\bar u p\cdot x} \bb{\phi _{\mathcal P} (u) + \frac{m _{\mathcal P}^2 x^2}{16} \hat A (u)} \nn - \frac i2 f _{\mathcal P} m _{\mathcal P}^2 \frac{x _\mu}{p\cdot x} \int _0^1 du\ e^{i\bar u p\cdot x} \hat B (u) \label{12}\\
	\langle \mathcal P(p) \vert \bar u (x) \sigma _{\xi h}\gamma _5 d(0) \vert 0 \rangle &= \frac{i}{6} \mu _{\mathcal P} (1-\tilde \mu _{\mathcal P}^2) (p _\xi x _h - p _h x _\xi) \int _0 ^1 du \ e^{i\bar u p \cdot x} \phi _\sigma (u) \label{13}\\
	\langle \mathcal P(p) \vert \bar u (x) \gamma _\mu g _s  G _{\alpha\beta} (ux) d(0) \vert 0 \rangle &= i p _\mu (p _\alpha x _\beta - p _\beta x _\alpha) \frac{1}{p\cdot x} f _{\mathcal P} m _{\mathcal P}^2 \int \mathcal D \alpha _i \ e^{i(\alpha _1 + u \alpha _3)p\cdot x} \mathcal V _\parallel (\alpha _i) \nn + i \Big\{ p _\beta \bb{g _{\mu\alpha} - \frac{1}{p \cdot x} (p _\mu x _\alpha + p _\alpha x _\mu)} \nn - p _\alpha \bb{g _{\mu\beta} - \frac{1}{p\cdot x} (p _\mu x _\beta + p _\beta x _\mu)} \Big\} \nn\times  f _{\mathcal P} m _{\mathcal P}^2 \int \mathcal D \alpha _i \ e^{i(\alpha _1 + u \alpha _3)q\cdot x} \mathcal V _\perp (\alpha _i) \label{14}\\
	\langle \mathcal P (p) \vert \bar u (x) \gamma _\mu \gamma _5 g _s G _{\alpha \beta} (ux) d(0) \vert 0 \rangle &= p _\mu (p _\alpha x _\beta - p _\beta x _\alpha) \frac{1}{p\cdot x} f _{\mathcal P} m _{\mathcal P}^2 \int \mathcal D \alpha _i \ e ^{i(\alpha 1 + u \alpha 3) p \cdot x} \mathcal A _\parallel (\alpha _i) \nn + \Big\{ p _\beta \bb{g _{\mu\alpha} - \frac{1}{p\cdot x} (p _\mu x _\alpha + p _\alpha x _\mu)} \nn - p _\alpha \bb{g _{\mu\beta} - \frac{1}{p\cdot x} (p _\mu x _\beta + p _\beta x _\mu )} \Big\} \nn \times f _{\mathcal P} m _{\mathcal P}^2 \int \mathcal D \alpha _i \ e^{i(\alpha _1 + u \alpha _3)p\cdot x} \mathcal A _\perp (\alpha _i) \label{15}\\
	\langle \mathcal P (p) \vert \bar u (x) \sigma _{\xi h} \gamma _5 g _s G _{\lambda\tau}(ux) d(0) \vert 0 \rangle &= i \mu _{\mathcal P} \Big\{ 
		p _\lambda p _\xi \bb{g _{h\tau} - \frac{1}{p\cdot x} \bb{p _h x _\tau + p _\tau x _h}}
		\nn - p _\lambda p _h \bb{g _{\xi \tau} - \frac{1}{p\cdot x} \pp{p _\xi x _\tau + p _\tau x _\xi}}
		\nn - p _\tau p _\xi \bb{g _{h\lambda} - \frac{1}{p\cdot x} \pp{p _h x _\lambda + p _\lambda x _h}}
		\nn + p _\tau p _h \bb{g _{\xi \lambda} - \frac{1}{p\cdot x} \pp{p _\xi x _\lambda + p _\lambda x _\xi}}
	\Big\} \nn \times \int \mathcal D \alpha _i \ e^{i[p+(\alpha _1 + u \alpha _3)q]\cdot x}\mathcal T (\alpha _i) \label{16}
\end{align}
where $ \tilde G _{\alpha\beta} = \frac 12 \varepsilon _{\alpha\beta \mu\nu} G ^{\mu\nu} $ is the dual gluon field strength tensor and $ \int \mathcal D \alpha _i = \int d\alpha _1 \ d\alpha _2 \ d\alpha _3 \ \delta(1-\alpha _1 - \alpha _2 - \alpha _3) $. We give the explicit forms of the only the relevant DAs in Section \ref{sec:3}. Performing a Fourier transformation on the QCD part and a Borel transformation on both sides of the correlation function with respect to the variables $ -{p'}^2 $ and choosing the coefficients of the structures $ P_\mu P _\nu P _\lambda $ and $ q _\mu q _\nu P _\lambda $, we obtain the following sum rules for the relevant form-factors $ F _0(q^2) $ and $ F _1 (q^2) $:
\begin{align}
F_0^q (Q^2) &= - \frac{1}{2} e^{m _{\mathcal P}^2/M^2} (\mathcal I _1 [\bar u^2, \phi _{\mathcal P}(u),1]
+ \mathcal I _2[u^2 \phi _{\mathcal P}(u), 1])\\
F_1^q (Q^2) &= - \frac 12 e^{m _{\mathcal P}^2/M^2} (\mathcal I _1 [u(1+u) \phi _{\mathcal P}(u), 1] 
+ \mathcal I _2[\bar u (1+\bar u) \phi _{\mathcal P} (u), 1])
\end{align}
where $ Q^2 := -q^2 $ and the $ \mathcal I _k[f(u),n] $ are defined as
\begin{align}
\mathcal I _1 [f(u), n] &= (-1)^n \int _0 ^{u _{10}} du \ \frac{F _{1n}(u)}{(n-1)! (M^2)^{n-1}} e^{-s _1(u)/M^2} 
\nn - \Big[
\frac{(-1)^{n-1}}{(n-1)!} e^{-s _1(u)/M^2} \sum _{\ell = 1}^{n-1} \frac{1}{(M^2)^{n-\ell - 1}} \frac{1}{s _1' (u)} \Big(\frac{d}{du} \frac{1}{s _1'(u)}\Big) ^{\ell -1} F _{1n} (u) 
\Big] _{u=u _{10}} \label{I1}\\
\mathcal I _2 [f(u), n] &= (-1)^n \int _{u _{20}}^1 du \ \frac{F _{2n}(u)}{(n-1)! (M^2)^{n-1}} e^{-s _2(u)/M^2} 
\nn - \Big[
\frac{(-1)^{n-1}}{(n-1)!} e^{-s _2(u)/M^2} \sum _{\ell = 1}^{n-1} \frac{1}{(M^2)^{n-\ell - 1}} \frac{1}{s _2' (u)} \Big(\frac{d}{du} \frac{1}{s _2'(u)}\Big) ^{\ell -1} F _{2n} (u) 
\Big] _{u=u _{20}} \label{I11}
\end{align}
with
\begin{align}
	s _1 (u) &= m _{\mathcal P}^2 u - \frac{u}{\bar u} q^2\\
	s _2 (u) &= m _{\mathcal P}^2 \bar u - \frac{\bar u}{u} q^2
\end{align}
and
\begin{align}
	F _{1n} (u) &= \frac{f(u)}{\bar u^n}\\
	F _{2n} (u) &= \frac{f(u)}{u^n}
\end{align}
$ u _{10} $ is a solution of the equation $ m _{\mathcal P}^2 u - uq^2/\bar u = s _0 $ and $ u _{20} $ is a solution of the equation $ m _{\mathcal P}^2 \bar u - \bar u q^2 /u = s _0 $. To keep the expressions short, we do not present the terms proportional to $ m _{\mathcal P}^2 $ but in our numerical analysis, we take their contributions, as well.
\section{Numerical analysis}\label{sec:3}
\begin{figure}
	[H]\centering 
	\includegraphics[width=0.7\textwidth]{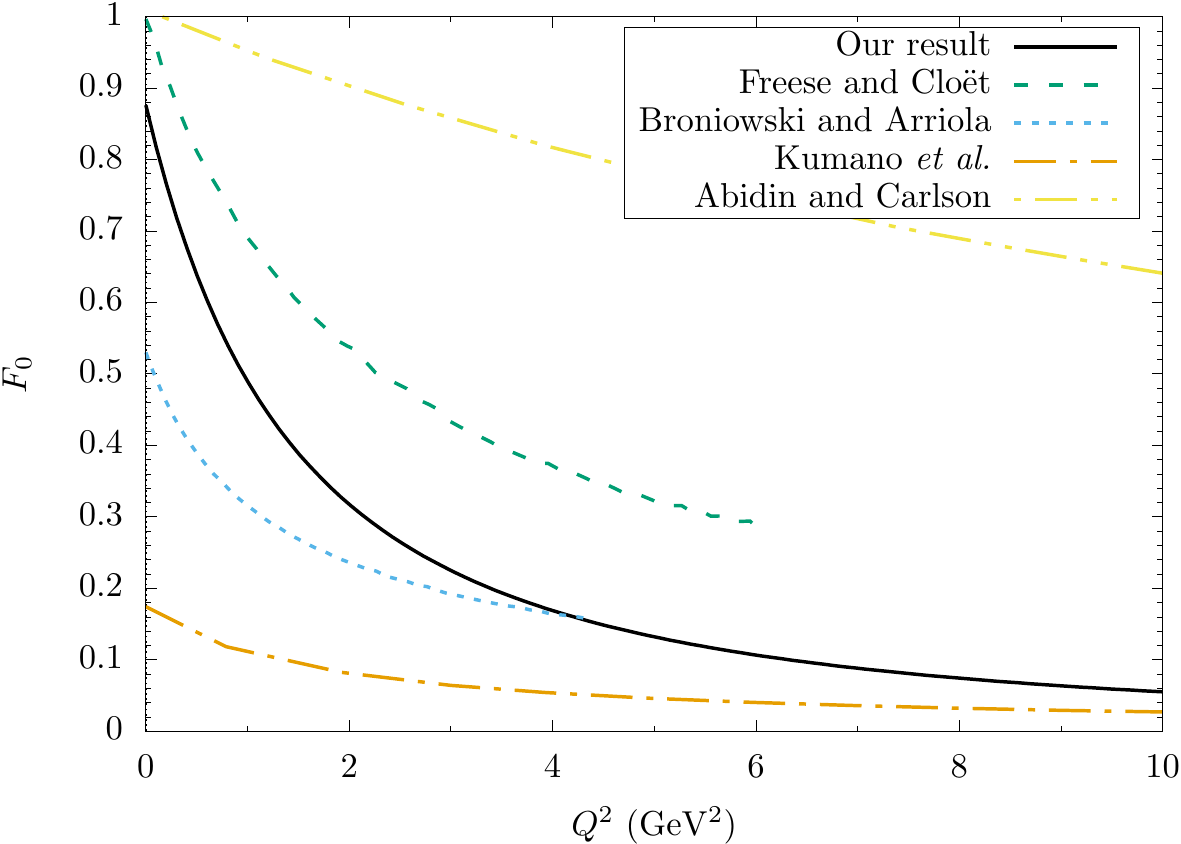}
	\caption{The GFFs of the $ \pi $ meson: $ F _0(Q^2) $ at $ s _0 = 0.7 {\rm\ GeV^2} $. Our results are compared to the ones in the chiral quark model \cite{ref:b2}, in the NJL model \cite{ref:10}, in the AdS/QCD theory \cite{ref:11}, and obtained by using the generalized distribution amplitudes \cite{ref:a2}.}
	\label{fig:1}
\end{figure}

\begin{figure}
	[H]\centering 
	\includegraphics[width=0.7\textwidth]{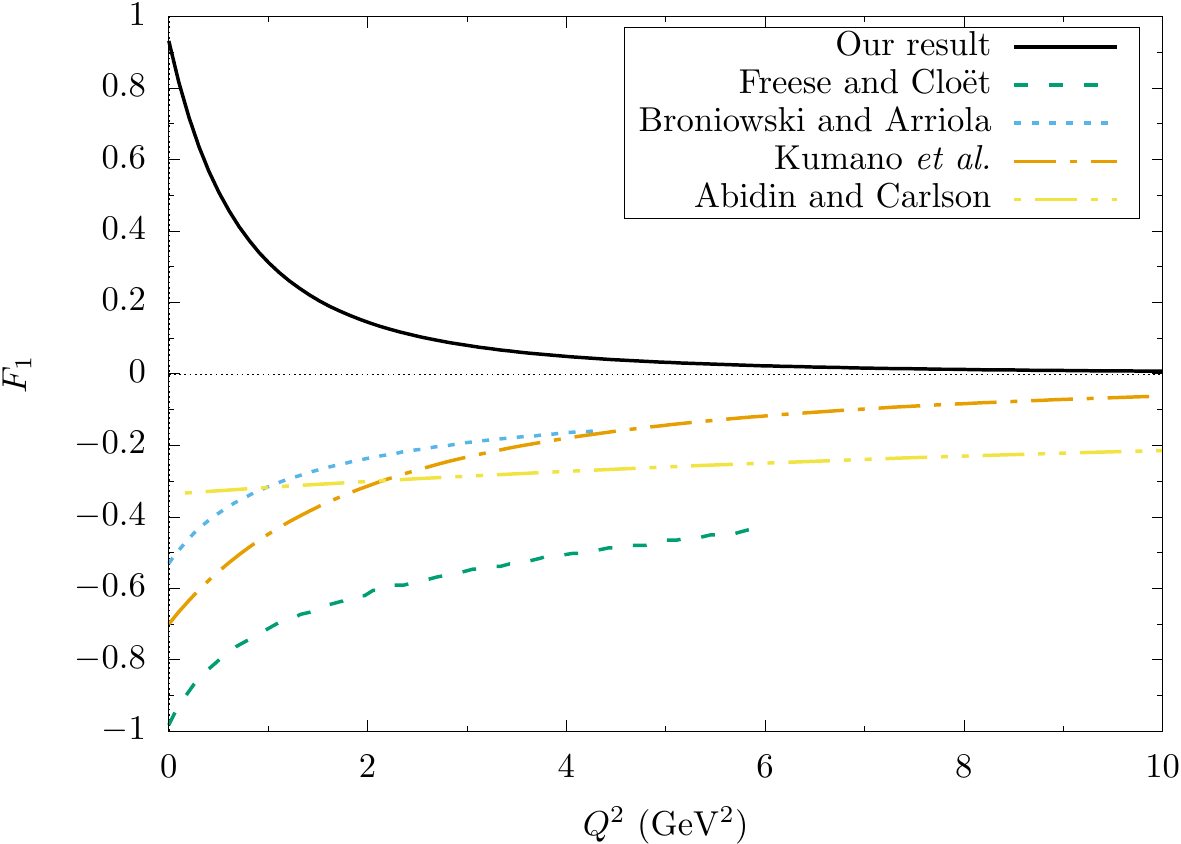}
	\caption{The same as Fig. \ref{fig:1} but for $ F _1(Q^2) $.}
	\label{fig:2}
\end{figure}

\begin{figure}
	[H]\centering 
	\includegraphics[width=0.7\textwidth]{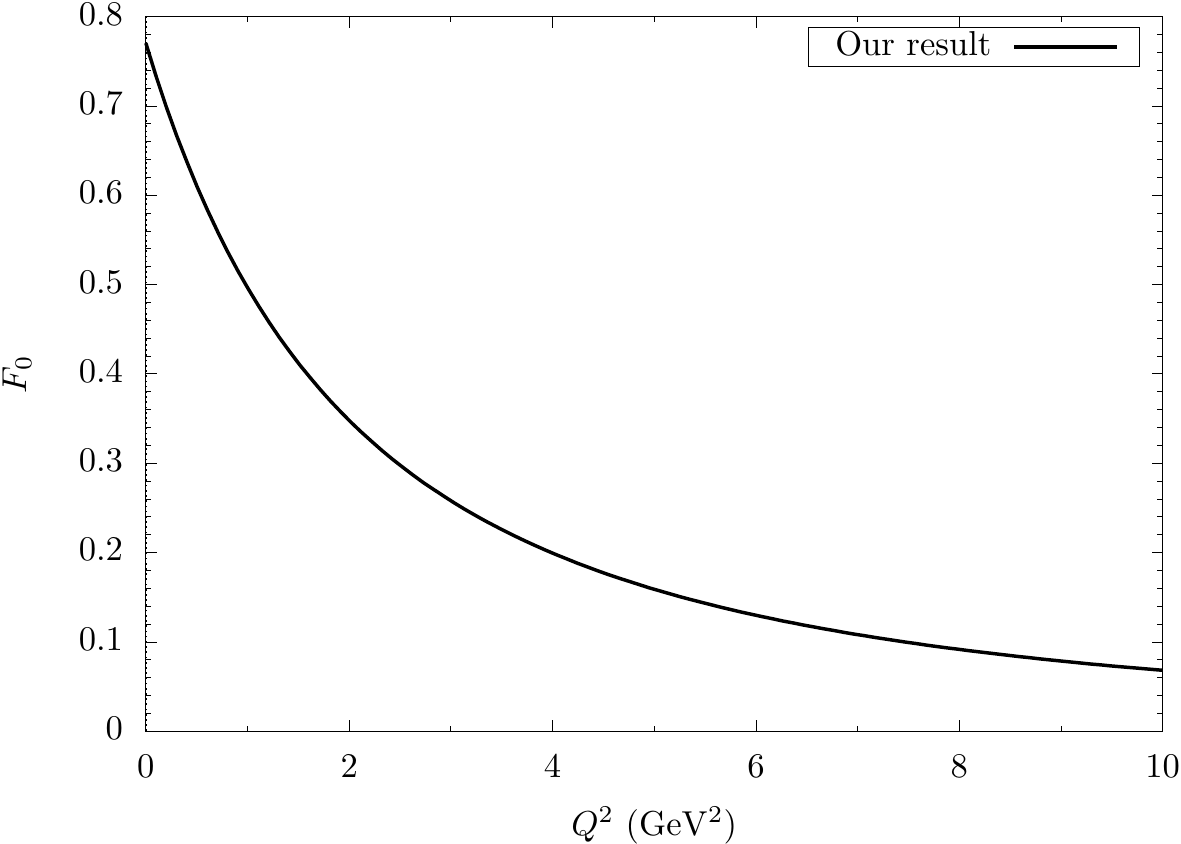}
	\caption{The GFFs of the $ K $ meson: $ F _0(Q^2) $ at $ s _0 = 1.05 {\rm\ GeV^2} $.}
	\label{fig:3}
\end{figure}

\begin{figure}
	[H]\centering 
	\includegraphics[width=0.7\textwidth]{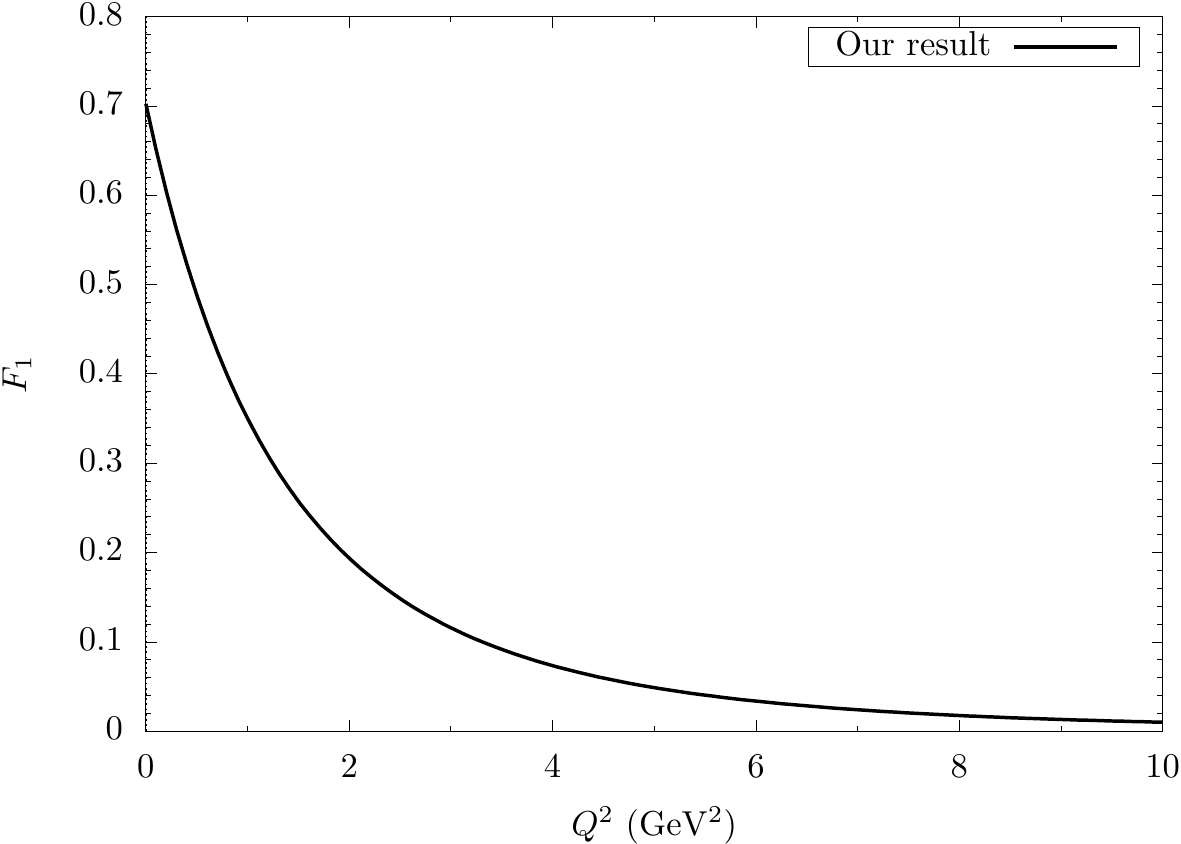}
	\caption{The GFFs of the $ K $ meson: $ F _1(Q^2) $ at $ s _0 = 1.05 {\rm\ GeV^2} $.}
	\label{fig:4}
\end{figure}

In this section, we present the numerical analysis of the LCSR for the GFFs of the $ \pi $ and $ K $ mesons by using Package X \cite{ref:20}. In the LCSR, we take the mass and the decay constant of the aforementioned pseudoscalar mesons to be $ m _\pi = 135 {\rm\ MeV} $, $ m _K = 497 {\rm\ MeV} $, $ f _\pi = 131 {\rm\ MeV} $, and $ f _K = 160 {\rm\ MeV} $, respectively. Another set of the fundamental input parameters of the LCSR is the pseudoscalar meson DAs of different twists. Even though the only DA that survives in the analytical results is $ \phi _{\mathcal P } (u) $, we would like to give all the relevant DAs \cite{ref:18,ref:19}:
\begin{align}
	\phi _{\mathcal P} (u) &= 6u \bar u \bb{1 + a _1^{\mathcal P} C_1(\xi ) + a _2^{\mathcal P} C _2^{3/2} (\xi)}\\
	\phi _P (u) &= 1 + \pp{30 \eta _3 - \frac{5}{2} \frac{1}{\mu _{\mathcal P}^2} } C _2^{1/2} (\xi) + \pp{-3\eta _3 w _3^{\mathcal P} - \frac{27}{20} \frac{1}{\mu _{\mathcal P}^2} - \frac{81}{10} \frac{1}{\mu _{\mathcal P}^2} a _2 ^\mathcal P} C _4^{1/2} (\xi)\\
	\phi _\sigma (u) &= 6u\bar u\bb{
		1 +
		\pp{5\eta _3 - \frac 12 \eta _3 w _3^{\mathcal P} - \frac{7}{20} \mu _{\mathcal P}^2 - \frac 35 \mu _{\mathcal P}^2 a _2 ^{\mathcal P}} C _2^{3/2} (\xi)
	}\\
	\mathcal T (\alpha _i) &= 360 \eta _3 \alpha _1 \alpha _2 \alpha _3 ^2 \bb{1 + w _3^{\mathcal P} \frac 12 (7\alpha _3 - 3)}
\end{align}
The $ C _n^k (x) $ are the Gegenbauer polynomials and $ \xi = u $ or $ \bar u $, depending on the position of the quarks in the matrix elements in Eqs. \eqref{11} -- \eqref{16}. The values of the parameters that appear in the DAs at the renormalization scale of $ \mu = 1 {\rm\ GeV} $ are as follows: $ a_1^\pi = 0 $, $ a _2^\pi = 0.44 $, $ a _1^K = 0.06 $, $ a _2^K = 0.25 $, $ \eta _3 = 0.015 $, $ w _3^\pi = -3 $, and $ w _3 ^K = -1.2 $.
\prg 
From the LCSR for the GFFs, one can see that in addition to the above-mentioned input parameters, they include two auxiliary parameters, i.e. the Borel mass parameter, $ M^2 $, and the continuum threshold, $ s _0 $. Physically measurable quantities should be independent of these auxiliary parameters. The working region of $ M^2 $ and the continuum threshold $ s _0 $ for the pseudoscalar mesons is determined from the analysis of two-point correlation function and the results are
\begin{align}
	0.6 {\rm\ GeV^2} < M^2 < 1.4 {\rm\ GeV^2}, \quad s _0 = 0.7 {\rm\ GeV^2} 
\end{align}
for the pion and 
\begin{align}
	1.0 {\rm\ GeV^2} < M^2 < 1.5 {\rm\ GeV^2}, \quad s _0 = 1.05 {\rm\ GeV^2}
\end{align}
for the kaon, respectively.
\prg 
The LCSR can give reliable predictions at sufficiently large negative values of $ Q^2 $ and we can determine the form-factors reliably for $ Q^2 \geq 1 {\rm\ GeV^2} $. However, one cannot apply the LCSR method for the values of $ Q^2 $ smaller than $ 1 {\rm\ GeV^2} $. By looking for a suitable fitting function, we may extend the results for the form-factors to the point $ Q^2 = 0 $ such that the fitting curve coincides with the LCSR predictions at $ Q ^2 \geq 1 {\rm\ GeV^2} $ region. Our numerical analysis reveals that the best fitting curve for the form-factors is given by
\begin{align}
	a \pp{1 + \frac{Q^2}{b}}^c 
\end{align}
We present the values of the parameters $ a $, $ b $, and $ c $ obtained for the GFFs of the pion and kaon in Table \ref{tab:1}. In Figs. \ref{fig:1} -- \ref{fig:4}, we plot the $ Q^2 $-dependence of each said form-factor. In Figs. \ref{fig:1} and \ref{fig:2}, we also present the $ Q^2 $-dependence of the form-factors $ F _0 (Q^2) $ and $ F _1(Q^2) $ obtained within the NJL model \cite{ref:10}, by using the generalized parton distribution \cite{ref:a2}, in the chiral quark model \cite{ref:b2}, and in the AdS/QCD theory \cite{ref:11}.
{\renewcommand{\arraystretch}{1.5}
\begin{table}
	[H]\centering 
	\caption{The fitting parameters for the GFFs of the pion and kaon.}
	\label{tab:1}
	\begin{tabular}
		{ccccc} 
		\hline 
		\hline 
		 &  \multicolumn{2}{c}{$ \pi $} &  \multicolumn{2}{c}{$ K $} \\
		\hline
		Parameter&  $ F _0 $ &  $ F _1 $ &  $ F _0 $ &  $ F _1 $\\
		\hline  
		 $ a $ & $ 0.876 $ & $ 0.932 $ & $ 0.770 $ & $ 0.703 $  \\
		 $ b $ & $ 2.457 $ & $ 1.951 $ & $ 3.798 $ & $ 5.198 $ \\
		 $ c $ & $ -1.704 $ & $ -2.636 $ & $ -1.879 $  & $ -3.957 $  \\
		\hline 
		\hline 
	\end{tabular}
\end{table}
}
By looking at Figs. \ref{fig:1} and \ref{fig:3}, recalling the fact that we consider only the quark part of the EMT, and in the light of Eq. \eqref{7}, we conclude that the quark contributions make up nearly 90\% (70\%) of the form-factor $ F _0 $ for the pion (kaon). Examining Figs. \ref{fig:2} and \ref{fig:4} and referring to Eq. \eqref{8}, one can see that the gluon contributions to the form-factor $ F _1 $ should be large and negative.  
\prg  
At the end of this section, we calculate the mean squared mass radii for $ \pi $ and $ K $ mesons in the light-cone frame. It is determined with the help of $ F _0 $ as 
\begin{align}
	\langle r ^2 \rangle _{\rm LC} = - 4 \frac{d F _0 (Q^2)}{dQ^2} \Big | _{Q^2=0} 
\end{align}
Using the $ Q^2 $-dependence of the $ F _0 $ form-factor for the pion and kaon, for mean squared mass radii we obtain
\begin{align}
	\sqrt{\langle r _\pi ^2 \rangle} &= 0.31 {\rm\ fm} \\
	\sqrt{\langle r _K ^2 \rangle} &= 0.24 {\rm\ fm} 
\end{align}
For comparison, we present the results of the works where $ \langle r _\pi ^2 \rangle $ is calculated within the NJL model, the AdS/QCD theory, and in different models \cite{ref:a1}. The results are 
\begin{align}
	\sqrt{\langle r _\pi^2 \rangle } = \begin{cases}
	0.27 {\rm\ fm}\ \mbox{\cite{ref:10}}\\
	0.29 {\rm\ fm} \ \mbox{\cite{ref:8}} \\
	0.34 {\rm\ fm} \ \mbox{\cite{ref:a1}}
	\end{cases}
\end{align}
The empirical value for $ \sqrt{\langle r _\pi^2 \rangle} $ extracted from KEKB data is between 0.26 and 0.32 fm \cite{ref:a2}. From the presented results for $ \sqrt{\langle r _\pi^2 \rangle} $, we find that our result is in good agreement with the predictions of other approaches.
\prg 
We also see that our result on mass radius is smaller than the charge radius which is equal to 0.51 fm for the pion \cite{ref:a3}.
\section{Conclusion}\label{sec:4}
In the present work, we calculate the GFFs appearing in the matrix element of the EMT between light pseudoscalar mesons, the pion and kaon states, within the LCSR. Furthermore, we calculate the mean squared mass radii of these mesons and compare our results with predictions of NJL, AdS/QCD, and other approaches. Our results for the pion mean squared mass radius is in good agreement with the predictions of other approaches. We also obtain that the mean square mass radius is smaller than the charge radius for the pion. 
%\appendix
%\section{}\label{app:A}
%\section{}\label{app:B}
\bibliography{paper}
\newpage 
\end{document}